\documentclass[reprint, amsmath,amssymb, aps,]{revtex4-2}
\usepackage{graphicx}% Include figure files
\usepackage{dcolumn}% Align table columns on decimal point
\usepackage{bm}% bold math
\usepackage{graphicx}
\usepackage{here}
\usepackage{color}

\begin{document}
\preprint{APS/123-QED}

\title{Efficient NMR measurement and data analysis supported by the Bayesian inference : The case of the heavy fermion compound YbCo$_2$Zn$_{20}$}% Force line breaks with \\
\author{H. Ueda, S. Katakami, M. Okada}
\email{okada@edu.k.u-tokyo.ac.jp}
\affiliation{
 Department of Complexity Science and Engineering, Graduate School of Frontier Sciences, The University of Tokyo, Kashiwa, Chiba 277-8561, Japan
}
\author{S. Yoshida, Y. Nakai, T. Mito}
\affiliation{
 Department of material Science, Graduate School of Science, University of Hyogo, Ako-gun 678-1297, Japan
}

\author{Masaichiro Mizumaki}
\affiliation{
    Faculty of Science, Course for Physical Sciences, Kumamoto University, Kurokami, Kumamoto 860-8555, Japan
}

\begin{abstract}
We propose a data-driven technique to infer microscopic physical quantities from nuclear magnetic resonance (NMR) spectra, in which the data size and quality required for the Bayesian inference are investigated.
The $^{59}$Co-NMR measurement of YbCo$_2$Zn$_{20}$ single crystal generates complex spectra with 28 peaks.
By exploiting the site symmetry in the crystal structure, the isotropic Knight shift $K_{\mathrm{iso}}$ and nuclear quadrupole resonance (NQR) frequency $\nu_Q$ were estimated to be $K_{\mathrm{iso}} = 0.7822 \pm 0.0090 \%, \nu_Q = 2.008 \pm 0.016$ MHz ($T = 20$ K and $H \simeq 10.2$ T) 
by analyzing only 30 data points from one spectrum.
The estimation of $\nu_Q$ is consistent with the precise value obtained in the NQR experiment. 
Our method can significantly reduce the measurement time and the computational cost of data analysis in  NMR experiments.
\end{abstract}

%\keywords{Suggested keywords}%Use showkeys class option if keyword
                              %display desired
\maketitle

%\tableofcontents

\section{Introduction}
Nuclear Magnetic Resonance (NMR) is an experimental method that observes the energy split of the nuclear spin under magnetic field to investigate the interaction between the nuclei and their environments.
In the field of condensed matter physics, various phenomena in solids, including magnetic order, superconducting transition, and structural transition have been studied by using the magnetic moment and the quadrupole moment of nuclei as probes(see for review~\cite{portis-magnetism,supercon,structure}).\par
When the NMR spectra have complex structures, the process of estimating physical quantities from the spectral data is a cumbersome task.
For example, $^{1}$H-NMR of organic molecules shows complicated split owing to the magnetic dipole-dipole interaction between $^{1}$H nuclei.
The efficient methods to infer from the NMR spectra the parameters of the effective model, Heisenberg model, have been studied~\cite{dashti_spin_2017,dashti_applications_2018,sels_quantum_2020}.
NMR measurements for nonorganic crystals can also produce convoluted spectra when the nuclear spin $I > 1$ and the nuclear quadrupole splitting occurs.
This is the case for our target material YbCo$_2$Zn$_{20}$~\cite{ishida_nmr_2012,mito_low-temperature_nodate}.\par

\begin{figure}[b]
    \begin{center}
    \includegraphics[width=0.8\columnwidth]{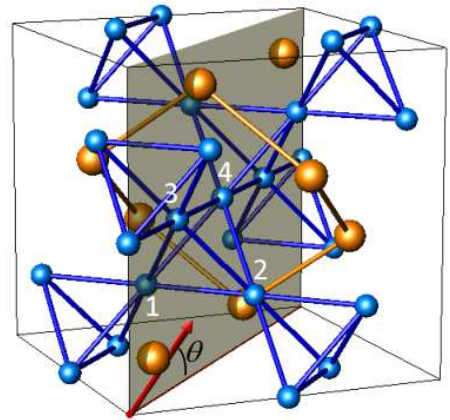}% Here is how to import EPS art
    \end{center}
    \caption{Schematic view of the crystal structure of YbCo$_2$Zn$_{20}$. The Zn atoms are omitted for clarity. 
    The arrow denotes the field direction when the magnetic field is rotated within the (1 $\bar{1}$ 0) plane.
    The indices of 1 to 4 label four Co ions forming a tetrahedron. }
\end{figure}

Yb-based compound YbCo$_2$Zn$_{20}$ shows extremely large specific heat ($C/T \simeq 7.9$/mol\ K$^2$) at low temperature~\cite{torikachvili_six_2007}, and undergoes nonmagnetic-magnetic transition under relatively low pressure $\sim$1 GPa~\cite{saiga_pressure-induced_2008}.
Strong field effects on YbCo$_2$Zn$_{20}$ have also been intensively studied, and field-induced Fermi liquid state and metamagnetic behavior have been discovered~\cite{ishida_nmr_2012,mito_mechanism_2012,takeuchi_metamagnetic_2011,ohya_strong_2010}.\par
As shown in Fig. 1, YbCo$_2$Zn$_{20}$ crystalizes in the space group $Fd\bar{3}m$ and $^{59}$Co sites form pyrochlore structure (site symmetry: $\bar{3}m$).
$^{59}$Co has nuclear spin $I = \frac{7}{2}$ and the four Co ions (site 1 to 4 in Fig. 1) forming a tetrahedron produces different NMR spectra under magnetic field, so that 28 peaks are expected to arise at maximum. 
{\it{Ishida et al.}} repeatedly rotated a single crystal and measured $^{59}$Co-NMR spectra to orient the magnetic field within the (1$\bar{1}$0) plane (gray plane in Fig. 1), where NMR peaks from the site 1 and 2 coincide.
They also determined the [001] direction and performed NMR measurements with the controlled angle $\theta$ in Fig. 1.
The physical quantities included in the model Hamiltonian of $^{59}$Co site (the model parameters), such as Knight shift,  were determined by manual fitting, and they investigated field and temperature dependence of the isotropic Knight shift. 
Figure 2 shows an example of manual fitting of NMR spectra. 

\begin{figure}[h]
    \begin{center}
    \includegraphics[width=0.9\columnwidth]{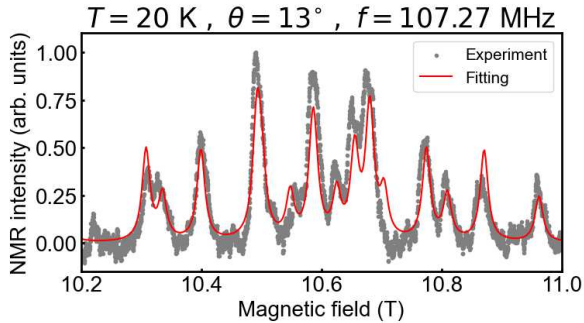}% Here is how to import EPS art
    \end{center}
    \caption{(Color online) Field swept $^{59}$ Co-NMR spectra of YbCo$_2$Zn$_{20}$ (temperature $T = 20$ K, $\theta = 13^{\circ}$, NMR frequency $f = 107.27$ MHz. Red line shows an example of manual fitting. )}
\end{figure}

In this study, we demonstrate that the measurement time and the cost of data analysis can be greatly reduced by applying the Bayesian framework to the estimation of Knight shift.
We will treat the direction of the crystal as a random variable so that the experimenters do not have to iteratively take NMR spectra to determine the precise angle of the sample.
While {\it{Ishida et al.}} used a value of the nuclear quadrupole resonance (NQR) frequency $\nu_Q$ directly obtained from a $^{59}$Co-NQR measurement~\cite{mito_low-temperature_nodate}, we only assume that $\nu_Q$ takes a typical value for $^{59}$ Co nuclei (about within 0 $\sim$ 15 MHz, the prior distribution of $\nu_Q$ will be given in Eq. (5)), and the precise value will be estimated by the Bayesian inference.
Of more than 5000 data points shown in Fig. 2, we will only use 30 points. \par
The rest of this paper will be organized as follows. 
Sec. II explains the forward model of the NMR spectra and the Bayesian framework we used.
Sec. III investigates the minimum data size and quality required for the estimation of physical parameters. 
The results obtained in this section are used to determined the number of data points we analyze in the next section.
Sec. IV describes the result of the analysis of real experimental data.
At last, Sec.V concludes this paper.
\section{Method}
\subsection{Model of the NMR spectra}
As we treat the orientation of the crystal as random variables, we will need two angle variables to specify the field direction. 
When the field direction is within the (1$\bar{1}$0) plane, the latitude of the field direction is $\theta$ and the longitude is $45^{\circ}$. 
We treat both of the latitude $\theta$ and the longitude $\phi$ as random variables.\par
To specify the nuclear Hamiltonian of $^{59}$Co sites, Kight shift tensor $\tilde{K}$ and the parameters related to the nuclear quadrupole interactions will be introduced, in addition to the magnitude $H_{\mathrm{ex}}$ and the direction $\theta, \phi$ of external field.
The axial symmetry at the $^{59}$Co site ($\bar{3}m$) leads to two independent tensor components $K_x$ and $K_z$. 
We can also deduce the asymmetry parameter $\eta = 0$ from the axial symmetry, so that the only unknown parameter with respect to the nuclear quadrupole interaction is $\nu_Q$.
The precise form of the model Hamiltonian will be explained in Appendix. A. \par
When $\bm{x} = (H_{\mathrm{ex}}, \theta, \phi, K_x, K_z, \nu_Q)$ is specified, the Hamiltonians for the site 1 to 4 $\mathcal{H}_1(\bm{x})\ ,..., \ \mathcal{H}_4(\bm{x})$ determine the 28 sets of resonance frequency and transition probability $\{(\mu_i, a_i)\}_{i = 0}^{27}$. 
In frequency-swept spectral measurements where the magnetic field is fixed to $H_{\mathrm{ex}}$, the NMR signals $y_0,...,y_{N-1}$ with respect to measurement points $f = f_0,...,f_{N-1}$ can be modeled as Eq. (1).
\begin{equation}
    y_n = A \sum_{i=0}^{27} \frac{\beta}{\pi} \frac{a_i}{(f_n - \mu_i)^2 + \beta^2} + \epsilon_n
\end{equation}
Here, $A$ is the ratio between the signal intensity and transition probability, and $\beta$ denotes the line width. 
$\epsilon_0,...,\epsilon_{N-1}$ are independent gaussian noise with their mean 0 and standard deviation $\sigma$.
Note that, when the frequency range is not sufficiently narrow, it is not a good approximation to treat $A$ as a constant, and the frequency dependence of $A$ should be introduced. 
This is due, for example, to the $f$ dependence of the tuning and matching conditions of the NMR probe's tank circuit, and to the $f$ variation of the signal intensity obtained with the standard spin-echo method we employed.
For metals, the penetration depth of the sample can also affect the $f$ dependence of $A$.
In such cases, a model for $A(f)$ should be constructed.\par
For field-swept NMR where the NMR frequency is fixed to $f$, the set of resonant frequency and transition probability $\{(\mu_i(H_n), a_i(H_n))\}_{i = 0}^{27}$ is obtained for each of the magnetic field $H_{\mathrm{ex}} = H_0,...,H_{N-1}$.
(The arguments included in $\mu_i$ and $a_i$ other than the magnetic field, such as $\theta, \phi, \tilde{K}$ and $\nu_Q$, are omitted for clarity.)
The NMR signal $y_0,...,y_{N-1}$ is modeled by Eq. (2).
\begin{eqnarray}
    y_n &=& A \sum_{i=0}^{27} \frac{\beta}{\pi} \frac{a_i(H_n)}{(f - \mu_i(H_n))^2 + \beta^2} + \epsilon_n\\\nonumber
        &=& y(H_n, \Theta) + \epsilon_n
\end{eqnarray}
where $\Theta = \{ A,\beta, \nu_Q, K_x, K_z, \theta, \phi \}$ is the set of parameters to be estimated. This paper describes the results for the field-swept NMR.\par
Note that the Bayesian approach can be applied to more general use in condensed matter physics, such as the analysis of the NMR spectra of samples with phase transitions.
For example, when the target material undergoes a first order phase transition, the NMR spectra should be modeled as the summation of NMR spectra from the two phases.
One may construct a model for such NMR spectra, and estimate the Knight shift for each domain.
The Bayesian method can also be used to detect the phase transition itself, by calculating the Bayes' free energy defined later. 
The detection of phase transition can be done by comparing the two models : one model which is the summation of two NMR spectra from two phases, and the other model which assumes that there exists only one domain in the sample.
If the Bayes' free energy of the former model is less than that of the latter model, it means that the model with two phases is more plausible, which is a data-driven detection of the phase transition.
\subsection{Bayesian framework}
For each of the parameters included in $\Theta$, we assume independent prior distribution as written below.
\begin{eqnarray}
    A &\sim& \gamma(\lambda_A,\eta_A)\\
    \beta &\sim& \gamma(\lambda_{\beta},\eta_{\beta})\\
    \nu_Q &\sim& \gamma(\lambda_{Q},\eta_{Q})\\
    K_x, K_z &\sim& \mathcal{N}(0, \Delta K)\\
    \theta &\sim& \mathcal{N}(\theta_0, \Delta \theta)\\
    \phi &\sim& \mathcal{N}(\phi_0, \Delta \phi)
\end{eqnarray}
Here, $\mathcal{N}(a,b)$ denotes a gaussian distribution with its mean $a$ and standard deviation $b$. $\gamma(\lambda, \eta)$ is a gamma distribution with its probability density function described in Eq. (9).
\begin{equation}
    p(x| \lambda, \eta) = \frac{1}{\Gamma(\eta)} \lambda^{\eta} x^{\eta-1} \exp (-\lambda x)
\end{equation}
The prior distribution of $\Theta$ defined above is denoted by $p(\Theta)$.\par
For both numerical experiments and real data analysis, we use these parameters for the prior distribution.
$\lambda_A = \lambda_{\beta} = 10,\ \eta_A = \eta_{\beta} = 2,\ \lambda_Q = 1,\ \eta_Q = 4,\ \Delta K = 5 \%,\ \Delta \theta = \Delta \phi = 5^{\circ}$.
$\theta_0$ and $\phi_0$ is the angle of the crystal roughly estimated from the direction of crystal face. $\Delta \theta = \Delta \phi = 5^{\circ}$ means this estimation has an error of $\sim 5^{\circ}$.
When one is given the dataset $\mathcal{D} = \{H_n, y_n\}_{n=0}^{N-1}$, the posterior distribution $p(\Theta | \mathcal{D})$ is given by the Bayes' theorem.
\begin{equation}
    p(\Theta | \mathcal{D}) = \frac{1}{Z(\sigma)} \exp (-\frac{N}{\sigma^2} E(\Theta)) p(\Theta)
\end{equation}
where $E(\Theta)$ is the energy function defined as below.
\begin{equation}
    E(\Theta) = \frac{1}{2N} \sum_{n=0}^{N-1} (y_n - y(H_n, \Theta))^2
\end{equation}
$Z(\sigma)$ is the normalization constant, and $f(\sigma) = -\log Z(\sigma)$ is called the Bayes' free energy.
If the noise level $\sigma$ is unknown, the minimizer of $f(\sigma)$ is the estimated value of $\sigma$~\cite{tokuda_simultaneous_2017,kashiwamura_bayesian_2022,moriguchi_bayesian_2022,tamura_data-driven_2020}.\par
To sample from the posterior distribution given by Eq. (10) and estimate the Bayes' free energy $f(\sigma)$, Markov's Chain Monte Carlo method is a fundamental tool.
However, the local minima of the energy function $E(\Theta)$ causes the multimodal nature of the posterior distribution, which makes the relaxation of the Monte Carlo method and the search for the MAP solution very slow.
The Exchange Monte Carlo method~\cite{hukushima_exchange_1996}, which was introduced in the study of spin-glass simulations, is also known to be effective for the Bayes' inference with multimodal posterior distributions~\cite{nagata_bayesian_2012} and widely used in the field of data-driven science~\cite{kashiwamura_bayesian_2022,moriguchi_bayesian_2022,katakami_bayesian_2022}.
We used this technique to perform the Bayes' inference of this study. The algorithmic details will be explained in Appendix. B.

\section{Limits of Hamiltonian inference}
{\it{Tokuda et.al}}~\cite{tokuda_intrinsic_2022} discussed the critical data size and quality required for the signal detection and the model selection when the data size and the signal noise ratio of spectral data are not sufficiciently large.
It is nice to know such limits when arranging costly experiments.\par
In this section, we perform numerical experiments to investigate the data size and accuracy of NMR spectra needed for the successful estimation of physical parameters such as the Knight shift. The results obtained here will be used to determine the appropriate number of data points we will process in Sec. IV.\par
The true parameters of the generated NMR spectral data are as follows (also shown in Table. I) .
$A = 0.200\ ,\ \beta = 0.100$ MHz\ ,\ $\nu_Q = 2.004$ MHz\ ,\ $K_x = 1.020\ \%\ ,\  K_z = 0.310\ \%\ ,\ \theta = 15.00^{\circ}, \phi = 42.00^{\circ}$.
We assume that the angle of the crystal was roughly estimated to be $\theta_0 = 13^{\circ}, \phi_0 = 45^{\circ}$ and these values will be used as prior distribution parameters defined in Eq.(7-8).
The data points of the magnetic field $H_n \ (n = 0,1,...,N-1)$ is generated by $H_n = H_{\mathrm{min}} + (H_{\mathrm{max}}-H_{\mathrm{min}})\cdot \frac{n}{N-1}$.
Throughout this section, $H_{\mathrm{min}} = 10.2$ T\ ,\ $H_{\mathrm{max}} = 10.9$ T. The NMR frequency is set to $f = 107.27$ MHz.

\subsection{Example : $N = 30\ ,\  \sigma = 0.1$}
\begin{figure*}[tbh]
    \begin{center}
        \includegraphics[width=0.8\linewidth]{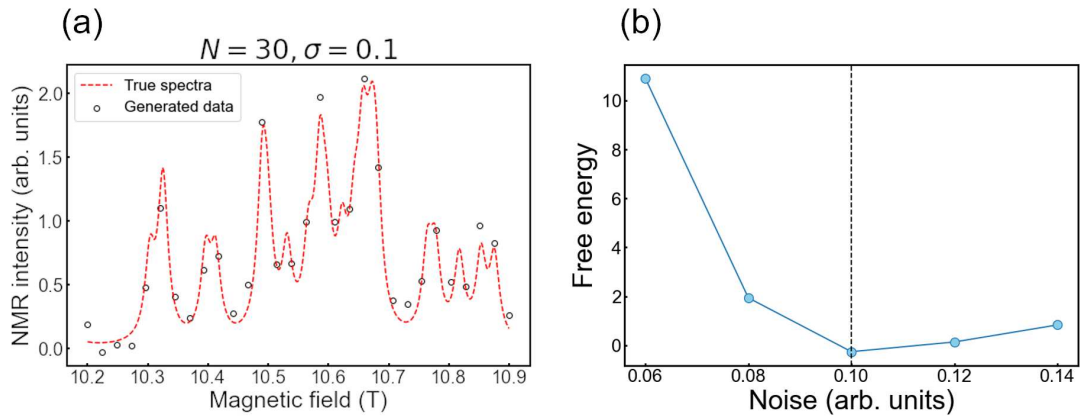}
    \end{center}
    \caption{(a) An instance of generated data. The dashed line shows the true spectra, and open dots are the data points with noise. (b) The result of noise estimation. The dashed line denotes the true standard deviation of the noise.}
\end{figure*}

\begin{figure*}[tbh]
    \begin{center}
        \includegraphics[width=\linewidth]{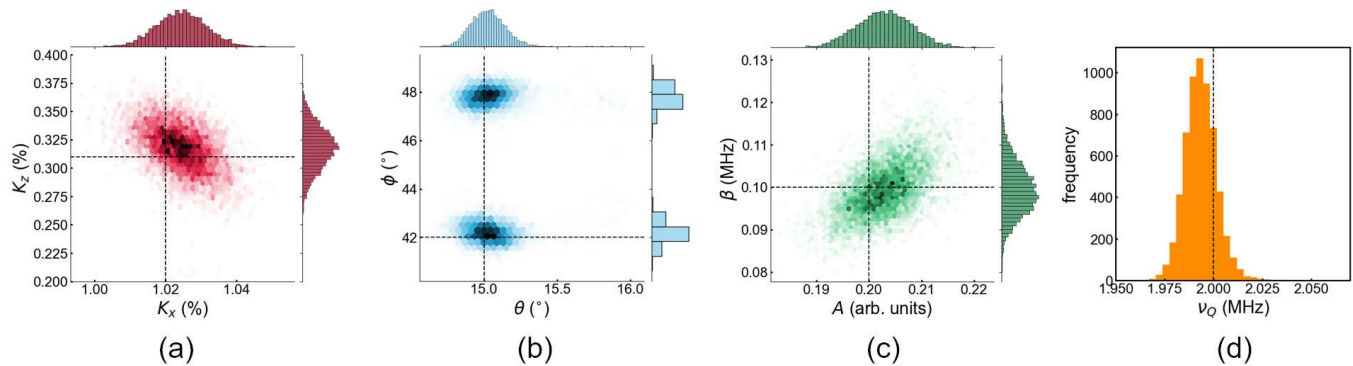}
    \end{center}
    \caption{The posterior distribution. (a)\ Knight shift $K_x, K_z$\ (b)\ Angle $\theta, \phi$\ (c)\ The relative signal amplitude $A$ and the line width $\beta$ (d) NQR frequency $\nu_Q$.}
\end{figure*}

First, we show the results for the case where the data size $N = 30$ and noise level $\sigma = 0.1$. Fig. 3 (a) is the true spectra and generated data points.
The standard deviation of noise $\sigma = 0.1$ is about $5\%$ of the maximum intensity of the signal.\par
If the noise level of the experimental data is unknown, we need a data-driven method to estimate $\sigma$.
We show the estimation of Bayes' free energy $f(\sigma)$ when $\sigma$ is varied in Fig. 3 (b). The noise level is successfully estimated to be $\sigma = 0.1$.
In this way, we can determine the noise from experimental data. In the following numerical experiments, we assume that the noise level $\sigma$ is already determined when we estimate physical parameters.

Figure 4 shows the posterior distribution of the parameters. The split of the posterior distribution of $\phi$ is the result of mirror symmetry about $(1\bar{1}0)$ plane.
The estimated value of each parameter is shown in Table I.
\begin{table}[tbh]
    \label{table:SpeedOfLight}
    \centering
     \begin{tabular}{clll}
      \hline
      Parameter & True value & Estimated value & Unit \\
      \hline \hline
      $K_x$ & 1.020 & 1.024 $\pm$ 0.014 & \% \\
      $K_z$ & 0.310 & 0.315 $\pm$ 0.040 & \% \\
      $\theta$ & 15.00 & 15.07 $\pm$ 0.26 & degree \\
      $\phi$ & 42.00 & 42.24 $\pm$ 0.73 & degree \\
      $A$ & 0.200 & 0.203 $\pm$ 0.010 & arb. units \\
      $\beta$ & 0.100 & 0.101 $\pm$ 0.012 & MHz \\
      $\nu_Q$ & 2.000 &1.994 $\pm$ 0.016 & MHz \\ 
      \hline
     \end{tabular}
     \caption{Parameter estimation for the case $N = 30, \sigma = 0.1$. The error bars represent the 95 \% confidence intervals. The estimated value and the error bar of $\phi$ were calculated using the half of the prior distribution where $\phi < 45^{\circ}$.}
\end{table}

The posterior distribution of the isotropic component of the Knight shift $K_{\mathrm{iso}} = \frac{2K_x + K_z}{3}$ can also be obtained. 
The estimated value is $K_{\mathrm{iso}} = 0.788 \pm 0.012 \ \%$.
The error bar of $K_{\mathrm{iso}}$ is relatively smaller compared with that of $K_x$ and $K_z$, which corresponds to the negative correlation between $K_x$ and $K_z$ seen in Fig. 4(a).

\subsection{Limits of data size and noise}
\begin{figure*}[tbh]
    \begin{center}
        \includegraphics[width=0.95\linewidth]{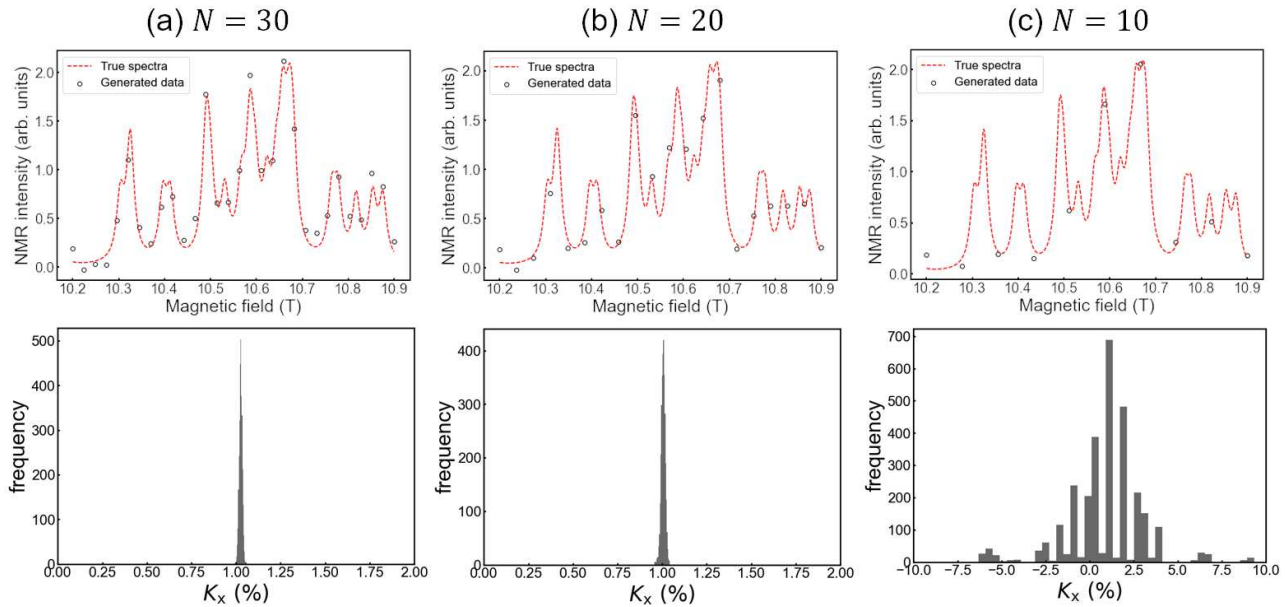}
    \end{center}
    \caption{Results of inference for the condition (a)$\ N = 30\ $(b)$\ N = 20\ $(c)$\ N = 10\ $. The circles in the upper panels represent the generated data. The lower row is the histogram of $K_x$ sampled from the posterior distribution.}
\end{figure*}

\begin{figure*}[tbh]
    \begin{center}
        \includegraphics[width=0.95\linewidth]{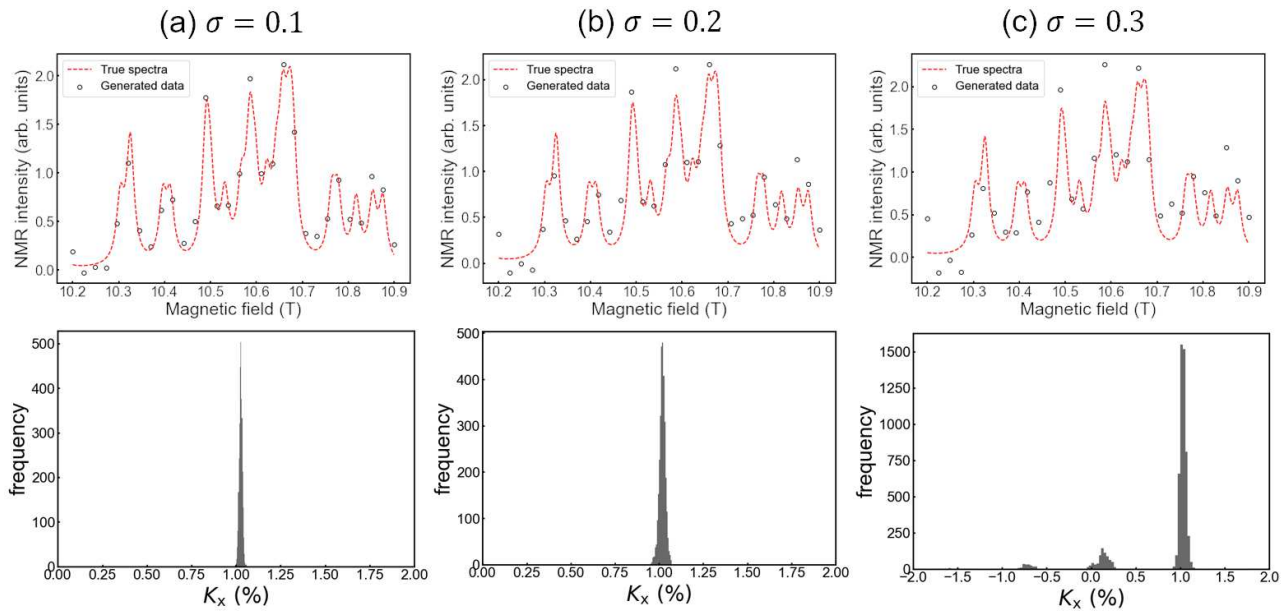}
    \end{center}
    \caption{Results of inference for the condition (a)$\ \sigma = 0.1\ $(b)$\ \sigma = 0.2\ $(c)$\ \sigma = 0.3\ $. The definition of upper and lower row is the same as Fig. 5.}
\end{figure*}

\begin{figure*}[tbh]
    \begin{center}
        \includegraphics[width=0.95\linewidth]{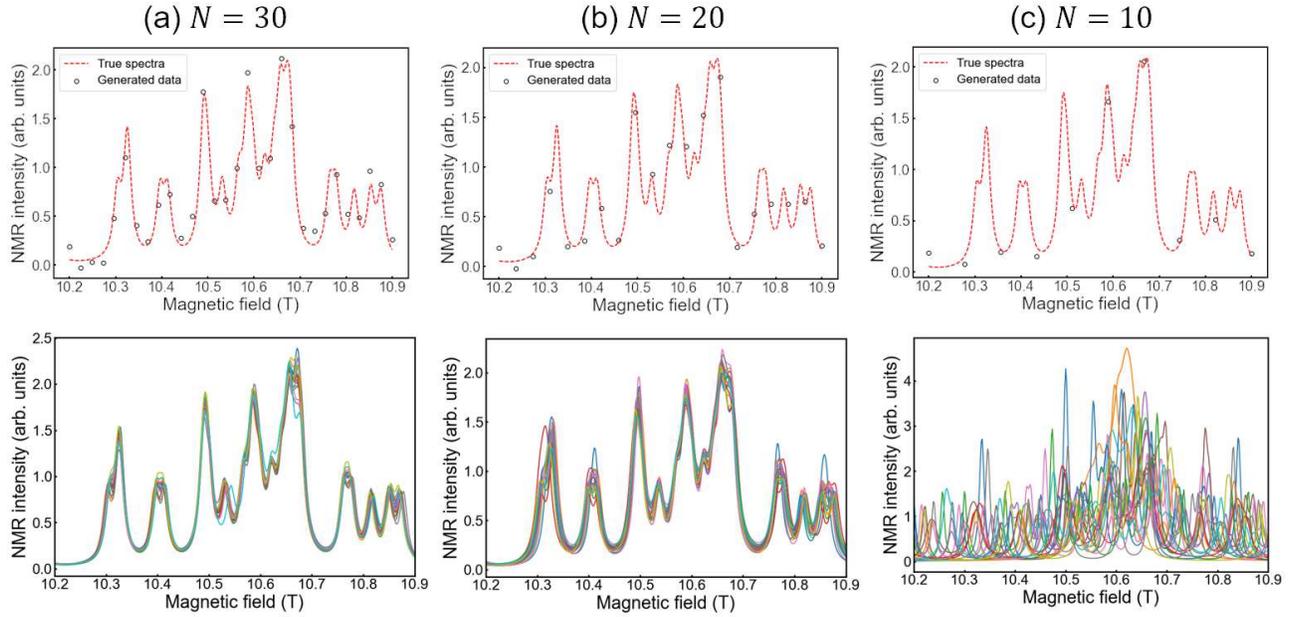}
    \end{center}
    \caption{The reconstruction of NMR spectra in the case (a)$\ N = 30\ $(b)$\ N = 20\ $(c)$\ N = 10\ $. The noise level is fixed to $\sigma = 0.1$. The circles in the upper panels represent the generated data. The lower row shows the 20 NMR spectra simulated by using 20 parameter sets sampled from the posterior distribution $p(\Theta | \mathcal{D})$ }
\end{figure*}

\begin{figure*}[tbh]
    \begin{center}
        \includegraphics[width=0.95\linewidth]{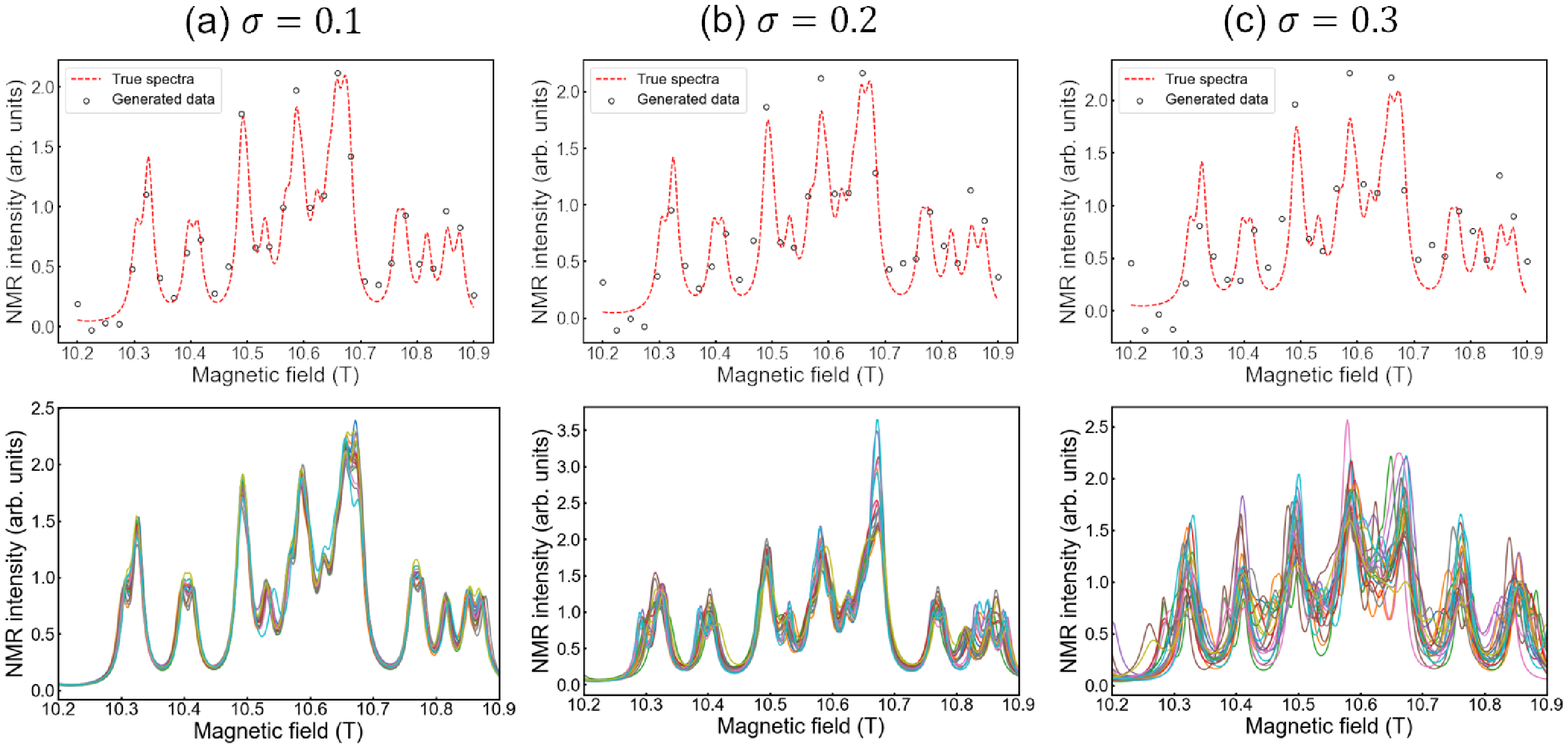}
    \end{center}
    \caption{The reconstruction of NMR spectra in the case (a)$\ \sigma = 0.1\ $(b)$\ \sigma = 0.2\ $(c)$\ \sigma = 0.3\ $. The data size is fixed to $N = 30$. The definition of the upper and lower row is the same as Fig. 7. }
\end{figure*}

Next, we show the results of the inference when the data size and data quality is reduced from the condition $N = 30, \sigma = 0.1$.
%In the case of the frequency-sweep NMR measurements, where the procedure of tuning and matching of the NMR probe is cumbersome, experimenters have to obtain physical information from limited number of data points.
%Even in the field-sweep NMR measurements, the experimental cost for each data point can be substantial if a weak signal is repeatedly accumulated while the external field is stopped.
%In these situations, the minimum data size required for the purpose of each experiment is helpful information for scientists.\par
The estimation of $K_x$ for the case $N = 10, 20$, and $30$ are shown in Fig. 5. In the condition $N = 20$, the confidence interval of $K_x$ is broadened to $K_x = 1.031 \pm 0.023 \ \%$ from $K_x = 1.024 \pm 0.014 \ \%$ of the case $N = 30$.
As to the data size $N = 10$, the posterior distribution of $K_x$ is dominated by its prior distribution.\par
We move on to the performance of parameter estimation when the noise level is increased.
%When the single crystal sample is small  or the NMR measurement is performed under low magnetic field, the signal-noise ratio deteriorates in general.
%In these situations, scientists can determine the accumulation time of NMR signal by the proposed method.\par
The results for noise level $\sigma = 0.1, 0.2$ and $0.3$ are shown in Fig. 6. In the condition $\sigma = 0.3$, the posterior distribution of $K_x$ is split to several peaks. \par
In the case of the frequency-sweep NMR measurements, where the procedure of tuning and matching of the NMR probe is cumbersome, experimenters have to obtain physical information from limited number of data points.
Even in the field-sweep NMR measurements, the experimental cost for each data point can be substantial if a weak signal is repeatedly accumulated while the external field is stopped.
In these situations, the results of the numerical experiment shown in this section can be used for planning the NMR experiment with the minimum cost.\par

It is illuminating to sample several parameter sets $\Theta$ independently from the posterior distribution, and plot the NMR spectra for those parameter sets as shown in Figs. 7-8. 
The performance of the reconstruction of the NMR spectra from the $N$ data points with noise $\sigma$ is visualized.
Moreover, the lower rows of Figs.7-8 tells us the optimal data point to add to these data. 
For example, in the case of $\sigma = 0.2$ in Fig. 8 , the uncertainty of the NMR signal is large around $10.67$ T. 
Thus, it is a wise decision to measure at this magnetic field and add this data to the previous $N$ data points.

\section{Analysis on experimental data}
{\it{Ishida et al.}} performed $^{59}$Co-NMR on YbCo$_2$Zn$_{20}$ single crystal and powder samples to study the field-induced effects~\cite{ishida_nmr_2012}.
Figure 9 shows the temperature and field dependence of $K_{\mathrm{iso}}$ which was determined by the model fitting to the NMR spectra.
The application of magnetic field suppresses the temperature dependent component of $K_{\mathrm{iso}}$ that arises from $4f$ moments, and they discussed that this is the sign of field-induced Fermi liquid state.
\begin{figure}[tbh]
    \begin{center}
        \includegraphics[width=0.8\linewidth]{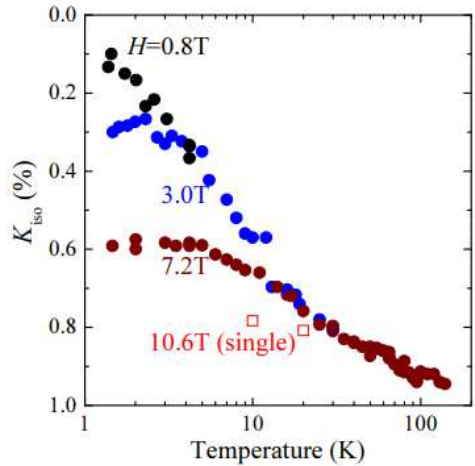}
    \end{center}
    \caption{Temperature and field dependence on $K_{\mathrm{iso}}$~\cite{ishida_nmr_2012}. The measurement on the condition $H \simeq$ 10.6 T was performed using single crystal, while the rest are done with powder sample.}
\end{figure}
\begin{figure}[tbh]
    \begin{center}
        \includegraphics[width=0.8\linewidth]{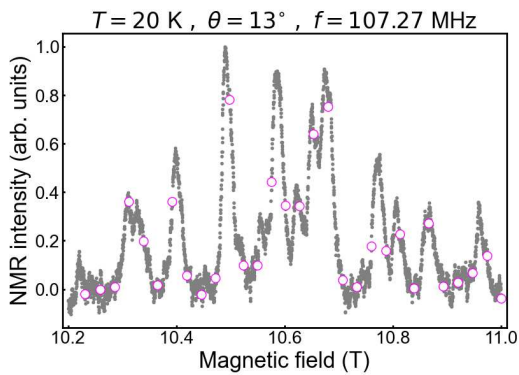}
    \end{center}
    \caption{The NMR spectra of YbCo$_2$Zn$_{20}$ single crystal, under the condition of temperature $T = 20$K, angle $\theta = 13^{\circ}$, and NMR frequency $f = 107.27$ MHz. The direction of the magnetic field is set within (1$\bar{1}$0) plane. The open points are the 30 data points we used for the Bayesian inference.}
\end{figure}
While $K_{\mathrm{iso}}$ was determined with manual fitting in their study, it is estimated with its uncertainty by the Bayesian inference in this study.
Our aim is to estimate $K_{\mathrm{iso}}$ with its 95 $\%$ confidence interval within $\pm 0.01 \%$.

\begin{figure}[tbh]
    \begin{center}
        \includegraphics[width=0.8\linewidth]{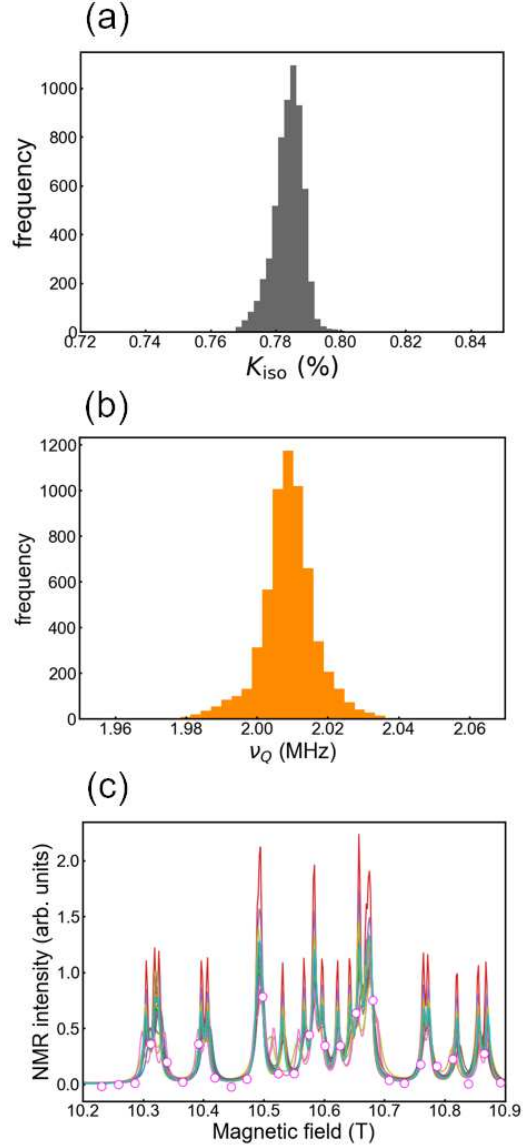}
    \end{center}
    \caption{The results of analysis on the experimental data. (a) The posterior distribution of $K_{\mathrm{iso}}$. The estimated value is $K_{\mathrm{iso}} = 0.7822 \pm 0.0090 \ \%$.
    (b) The posterior distribution of NQR frequency $\nu_Q$. The estimated value is $\nu_Q = 2.008 \pm 0.016$ MHz.
    (c) 20 spectra calculated from 20 parameter sets sampled from $p(\Theta|\mathcal{D})$. The open dots denote the experimental data at $T = 20$ K.
    }
\end{figure}
Figure 10 shows the experimental data we analyzed. The noise level is estimated as $\sigma = 0.05$, and the signal noise ratio is almost the same as that of the generated data shown in Fig. 3. Thus, considering the results shown in Figs. 5-6, it is expected that 30 data points would be enough data size for the estimation of $K_{\mathrm{iso}}$ with its 95 $\%$ confidence interval within $\sim \pm$0.01 $\%$. 
We extracted 30 points from more than 5000 data points as shown in Fig. 10, and estimated physical parameters by using these points. 
The posterior distributions of $K_{\mathrm{iso}}$ and $\nu_Q$ are shown in Fig. 11.(a) and (b), respectively.
We also show the NMR spectra calculated by using 20 parameter sets sampled from the posterior distribution in Fig. 11(c).

The estimated value of NQR frequency $\nu_Q = 2.008 \pm 0.016$ MHz is consistent with the precise estimation $\nu_Q = 2.004 \pm 0.001$ MHz obtained in the NQR experiment~\cite{mito_low-temperature_nodate}.
$K_{\mathrm{iso}} = 0.7822 \pm 0.0090 \ \%$ adds an uncertainty estimation to the analysis in ~\cite{ishida_nmr_2012}, and the error bar is sufficiently narrow to discuss the temperature and field dependence of $K_{\mathrm{iso}}$ shown in Fig. 9.

\section{Conclusion}
This study estimated physical parameters such as $K_{\mathrm{iso}}$ and $\nu_Q$ from NMR spectra with limited data size and signal noise ratio by applying the Bayesian framework, and demonstrated that the measurement time and the cost of data analysis can be greatly reduced. 
By the proposed method, experimenters can arrange NMR experiments by estimating the data size required for the accuracy of parameter estimation they need.

\appendix
\section{Model Hamiltonian}
Our model Hamiltonian of a $^{59}$Co site consists of the Zeeman term $\mathcal{H}_{\mathrm{Z}}$, the hyperfine interaction $\mathcal{H}_{\mathrm{hf}}$, and the nuclear quadrupole interaction $\mathcal{H}_{Q}$. 
We define the $z$ axis of the coordinate system attached to a $^{59}$Co site as its rotoinversion axis of order 3, and $x , y$ axes are defined so that a right-handed system is formed.  
$\vartheta$ denotes the angle between $z$ axis and the external magnetic field $\bm{H}_{\mathrm{ex}}$. Note that $\vartheta$ is determined by $\theta, \phi$ and the site index $i = 1,2,3,4$.\par
The Zeeman term $\mathcal{H}_{\mathrm{Z}}$ is given by
\begin{equation}
    \mathcal{H}_{\mathrm{Z}} = -\gamma_N \hbar \hat{\bm{I}} \cdot \bm{H}_{\mathrm{ex}}
\end{equation}
where $\gamma_N$ denotes the nuclear gyromagnetic ratio of $^{59}$Co nuclei and $\hat{\bm{I}}$ is the nuclear spin operator. The hyperfine interaction $\mathcal{H}_{\mathrm{hf}}$ is described by the hyperfine field $\bm{H}_{\mathrm{in}}$ as follows;
\begin{equation}
    \mathcal{H}_{\mathrm{hf}} = -\gamma_N \hbar \hat{\bm{I}} \cdot \bm{H}_{\mathrm{in}}
\end{equation}
Within the linear response, $\mathcal{H}_{\mathrm{hf}}$ is described by the Knight shift tensor $\tilde{K}$.
\begin{equation}
    \bm{H}_{\mathrm{in}} = \tilde{K} \bm{H}_{\mathrm{ex}}
\end{equation} 
and the site symmetry of $\bar{3}m$ constrains the tensor components of $\tilde{K}$ to the following form. 
\begin{equation}
    \tilde{K} = 
    \begin{pmatrix}
        K_x & 0 & 0 \\
        0 & K_x & 0\\
        0 & 0 & K_z\\
    \end{pmatrix}
\end{equation}
We use this model of hyperfine interaction in the Bayesian inference and estimate the tensor components $K_x, K_z$.
As to the nuclear quadrupole interaction, the asymmetry parameter $\eta$ is zero owing to the site symmetry. By choosing the direction of $x$ axis properly, $\mathcal{H}_{Q}$ is given by
\begin{widetext}
\begin{equation}
    \mathcal{H}_Q = \frac{h \nu_Q}{6} \left[ \frac{3\cos^2 \vartheta - 1}{2}\{3I_z^2 - I(I+1)\} + 
    \frac{3\sin\vartheta \cos\vartheta}{2} \{I_z (I_+ + I_-) + (I_+ + I_-)I_z \}+
    \frac{3 \sin^2 \vartheta}{4} (I_+^2 + I_-^2)      \right]
\end{equation}
\end{widetext}
where $I_{\pm}$ is the creation and annihilation operator, and $\nu_Q$ is the NQR frequency. $\nu_Q$ can be precisely measured by the NQR measurement, which is performed under zero magnetic field. The total Hamiltonian for one site is $\mathcal{H} = \mathcal{H}_{\mathrm{Z}} + \mathcal{H}_{\mathrm{hf}} + \mathcal{H}_{Q}$.

\section{The Exchange Monte Carlo method}
The normalization constant in Eq.(10) is given by the integral
\begin{equation}
    Z = \int \exp (-\frac{N}{\sigma^2} E(\Theta)) p(\Theta) d\Theta 
\end{equation}
The argument $\sigma$ of $Z(\sigma)$ is omitted for clarity. The Exchange Monte Carlo method introduces the inverse temperature $b$ to Eq.(B1).
\begin{equation}
    Z(b) = \int \exp (-\frac{Nb}{\sigma^2} E(\Theta)) p(\Theta) d\Theta 
\end{equation}
Note that $Z(1) = Z$ and $Z(0) = 1$.We estimate $Z(1)$ by using a sequence of inverse temperatures $0 = b_1 < b_2 < \cdots < b_{L-1} < b_L = 1$.
\begin{eqnarray}
    Z(1) &=& \prod_{l = 1}^{L-1} \frac{Z(b_{l+1})}{Z(b_{l})} \\
         &=& \prod_{l = 1}^{L-1} \frac{ \int \exp (-\frac{Nb_{l+1}}{\sigma^2}E(\Theta))p(\Theta) d\Theta}{ \int \exp (-\frac{Nb_{l}}{\sigma^2}E(\Theta))p(\Theta) d\Theta}\\
         &=& \prod_{l = 1}^{L-1}\left\langle \exp\left(\frac{N(b_{l+1}-b_l)}{\sigma^2}E(\Theta)\right)\right\rangle _{b_l}
\end{eqnarray}
Here,$\langle g(\Theta) \rangle_{b}$ denotes the expectation value of $g(\Theta)$ when $\Theta$ is sampled from the distribution $p(\Theta | b) \propto \exp (-\frac{Nb}{\sigma^2}E(\Theta))p(\Theta)$.
We used EMC to sample from $p(\Theta | b)$.EMC algorithm samples $L$ parameter sets $\Theta_1,..,\Theta_L$  subject to the distributions $p(\Theta_1 | b_1),..,p(\Theta_L|b_L)$ in pararell by the iteration of the two updates written below.\\
1. Update each parameter set $\Theta_l$ subject to $p(\Theta_l|b_l)$ by using Metropolis algorithm.\\
2. Exchange the value of $\Theta_l$ and $\Theta_{l+1}$($l = 1,2,..,L-1$) by using the following probability $u$,
\begin{eqnarray}
    u &=& \min (1,v)\\
    v &=& \frac{p(\Theta_l | b_{l+1})p(\Theta_{l+1} | b_{l})}{p(\Theta_l | b_{l})p(\Theta_{l+1} | b_{l+1})}
\end{eqnarray}
$Z$ and Bayes' free energy can be estimated by calculating the expectation value in the Eq.(B5) with respect to each inverse temperature $b_l$ by using EMC.
 In addition, the histogram of $\Theta_L$ provides the posterior distribution $p(\Theta | \mathcal{D})$.\\

\bibliography{NMR}

\end{document}